\documentclass[12pt]{article}
\newcommand{\pp}{{=\!\!\!|}}
\usepackage{amsmath}
\usepackage{amscd}
\usepackage{amsfonts}
\usepackage{amsthm}

\def\IN{\mathbb{N}}

\def\IX{\mathbb{X}}
\def\IJ{\mathbb{J}}
\def\IK{\mathbb{K}}
\def\IP{\mathbb{P}}

\begin{document}
\newcommand{\eqn}[1]{eq.(\ref{#1})}
\renewcommand{\section}[1]{\addtocounter{section}{1}
\vspace{5mm} \par \noindent
  {\bf \thesection . #1}\setcounter{subsection}{0}
  \par
   \vspace{2mm} } 
\newcommand{\sectionsub}[1]{\addtocounter{section}{1}
\vspace{5mm} \par \noindent
  {\bf \thesection . #1}\setcounter{subsection}{0}\par}
\renewcommand{\subsection}[1]{\addtocounter{subsection}{1}
\vspace{2.5mm}\par\noindent {\em \thesubsection . #1}\par
 \vspace{0.5mm} }
\renewcommand{\thebibliography}[1]{ {\vspace{5mm}\par \noindent{\bf
References}\par \vspace{2mm}}
\list
 {\arabic{enumi}.}{\settowidth\labelwidth{[#1]}\leftmargin\labelwidth
 \advance\leftmargin\labelsep\addtolength{\topsep}{-4em}
 \usecounter{enumi}}
 \def\newblock{\hskip .11em plus .33em minus .07em}
 \sloppy\clubpenalty4000\widowpenalty4000
 \sfcode`\.=1000\relax \setlength{\itemsep}{-0.4em} }

\vspace{2cm}

\begin{center}
{\bf A NOTE ON $N=(2,2)$ SUPERFIELDS IN TWO DIMENSIONS}

\vspace{1.4cm}

JORIS MAES and
ALEXANDER SEVRIN 

\vspace{.1cm}

{\em Theoretische Natuurkunde, Vrije Universiteit Brussel} \\
{\em and}\\
{\em The International Solvay Institutes}\\
{\em Pleinlaan 2, B-1050 Brussel, Belgium} \\
\end{center}

\vspace{2mm}

\centerline{ABSTRACT}
\vspace{-1 mm}  
\begin{quote}\small
Motivated by the results in {\tt hep-th/0508228}, we perform a careful analysis of the allowed linear constraints on 
$N=(2,2)$ scalar superfields. We show that only chiral, twisted-chiral and semi-chiral superfields are possible. 
Various subtleties are discussed.
\end{quote}
\baselineskip18pt
\addtocounter{section}{1}
\noindent

\vspace{5mm}
Non-linear $ \sigma $-models in two dimensions with an $N=(2,2)$ supersymmetry play a central role in the
description of type II superstrings in the absence of R-R fluxes. The interest in these models was recently rekindled 
as well in the physics as in the mathematics community. For physicists, these models allow for the study of 
compactifications in the presence of non-trivial NS-NS fluxes while for mathematicians the models provide a concrete 
realization of generalized complex geometries. The ideal setting for studying them is provided by $N=(2,2)$ superspace 
where the whole (local) geometry gets encoded in a single scalar function, the Lagrange density. The $N=(2,2)$ 
superfields and as a direct consequence the geometry of the resulting $ \sigma $-model as well, are fully 
characterized by their constraints. The analysis of auxiliary field configurations made in 
\cite{Bredthauer:2005zx} raised the suspicion that more general $N=(2,2)$ superfields than those known up till now 
might exist. This point will be explored in the present note.

The target space geometry of a bosonic non-linear $ \sigma $-model in two 
dimensions\footnote{We consider here only 
$ \sigma $-models without boundaries.} is characterized by the metric $g_{ab}$ and a closed 3-form (the torsion) 
$T_{abc}$ on it. 
We denote the local coordinates on the target manifold by $X^a$. The indices $a$, $b$, 
... run from 1 to $D$, with $D$ 
the dimension of the target manifold. Such a model can be lifted to an $N=(1,1)$ 
supersymmetric $ \sigma $-model 
without any further restrictions on the geometry. 
However, passing from $N=(1,1)$ to $N=(2,2)$ supersymmetry introduces additional geometric structure 
\cite{Alvarez-Gaume:1981hm}--\cite{Howe:1985pm}. 

Indeed $N=(2,2)$ requires the existence of two $(1,1)$ tensors, $J_+^a{}_b(X)$ and 
$J_-^a{}_b(X)$, on the target manifold. {\em 
On-shell} closure of the $N=(2,2)$ supersymmetry algebra is realized provided both 
$J_+$ and $J_-$ are complex 
structures. I.e.~they square to $-{\bf 1}$, $J_\pm^a{}_cJ_\pm^c{}_b=- \delta ^a_b$ and 
their Nijenhuis 
tensors\footnote{Out of two $(1,1)$ tensors $R^a{}_b$ and $ S^a{}_b$, one constructs a 
$(1,2)$ tensor 
${\cal N}[R,S]^a{}_{bc}$, the Nijenhuis tensor, as ${\cal N}[R,S]^a{}_{bc}= 
R^a{}_dS^d{}_{[b,c]}+R^d{}_{[b}S^a{}_{c],d}+R\leftrightarrow S$.} vanish, 
$ {\cal N}{[}J_\pm,J_\pm{]}^a{}_{bc}=0$. One 
finds that the off-shell non-closing terms in the algebra are proportional to the 
commutator of the two complex 
structures, ${[}J_+,J_-{]}$. Decomposing the tangent target space as 
$\ker{[}J_+,J_-{]}\oplus \mbox{coker}{[}J_+,J_-{]}$, 
one anticipates that the description of $\ker{[}J_+,J_-{]}$ will be possible with the fields at hand while the 
description of $\mbox{coker}{[}J_+,J_-{]}$ will require the introduction of additional auxiliary fields. 

Realizing the $N=(2,2)$ supersymmetry algebra is obviously not sufficient, the $N=(1,1)$ supersymmetric non-linear $ 
\sigma $-model has to be invariant under the additional supersymmetry transformations as well. One finds that this is 
indeed so provided the metric is hermitean with respect to {\em both} complex structures\footnote{This implies the 
existence of two two-forms $ \omega^\pm _{ab}=-\omega^\pm _{ba}= g_{ac}J^c_\pm{}_b$. In general they are not closed. 
Using eq.~(\ref{covconst}), one shows that $ \omega ^\pm_{[ab,c]}=\mp 2 J_\pm^d{}_{[a}T_{bc]d}=\mp (2/3)J^d_\pm{}_a
J^e_\pm{}_bJ^f_\pm{}_cT_{def}$, where for the last step we used the fact that the Nijenhuis tensors vanish.},
\begin{eqnarray}
J_\pm^c{}_a\,J_\pm^d{}_b\,g_{cd}=g_{ab}\,.\label{hermit}
\end{eqnarray}
Furthermore, both complex structures have to be covariantly constant, 
\begin{eqnarray}
0=\nabla_c^\pm \,J_\pm^a{}_b\equiv
\partial _c\,J_\pm^a{}_b+\Gamma^a_{\pm dc}J_\pm^d{}_{b}- \Gamma^d_{\pm bc}J_\pm^a{}_d\,,\label{covconst}
\end{eqnarray}
with the connections $\Gamma_\pm$ given by,
\begin{eqnarray}
\Gamma^a_{\pm bc}\equiv  \left\{ {}^{\, a}_{bc} \right\}
\pm T^a{}_{bc}\,,
\end{eqnarray}
where the first term at the right hand side is the standard Christoffel symbol and $T^a{}_{bc}=g^{ad}\,T_{dbc}$.

Finding an off-shell description of a general $N=(2,2)$ supersymmetric non-linear $ \sigma $-model remained for more 
than two decades an open problem. Recently this has been solved in \cite{Lindstrom:2005zr} (papers 
preparing the road to
this result are e.g. \cite{Buscher:1987uw}--\cite{Bogaerts:1999jc}). There it was shown that 
chiral, twisted-chiral \cite{Gates:1984nk} and semi-chiral \cite{Buscher:1987uw} multiplets are sufficient to describe 
any $N=(2,2)$ non-linear $ \sigma $-model. Roughly speaking one gets that when writing 
$\ker{[}J_+,J_-{]}=\ker(J_+-J_-)\oplus\ker(J_++J_-)$, 
$\ker(J_+-J_-)$ and $\ker(J_++J_-)$ resp.~can be integrated to chiral and twisted chiral multiplets 
resp.~\cite{Ivanov:1994ec}. Semi-chiral 
multiplets allow then for a description of $\mbox{coker}{[}J_+,J_-{]}$ \cite{Lindstrom:2005zr}, \cite{Sevrin:1996jr},
\cite{Bogaerts:1999jc}. 

However, chiral, twisted-chiral and semi-chiral multiplets are by no means the only representations of $d=2$, 
$N=(2,2)$ supersymmetry. Other representations are known such as linear \cite{Siegel:1979ai}--\cite{Deo:1985ix} 
and twisted linear multiplets \cite{Gates:1995du}. Having other 
multiplets at hand allow e.g.~for dual formulations of a model (see e.g.~\cite{Grisaru:1997ep} and references 
therein). In \cite{Bredthauer:2005zx}, a detailed analysis of potential auxiliary field configurations in $d=2$, 
$N=(2,2)$ $ \sigma $-models was performed. 
The resulting expressions were very involved and raise the pertinent question whether other solutions besides 
semi-chiral multiplets exist. This motivates the present note. We analyze constraints linear in derivatives on 
$N=(2,2)$ superfields in order to clarify this.

The $d=2$, $N=(1,1)$ superspace has two bosonic (lightcone) coordinates $ \sigma ^\pp\equiv \tau + \sigma $, 
$ \sigma ^=\equiv \tau - \sigma $ and two (chiral, real) fermionic coordinates $ \theta ^+$ and $\theta ^-$.
Passing to $N=(2,2)$ superspace requires the introduction of two additional (chiral, real) 
fermionic coordinates $ \hat\theta ^+$ 
and $\hat \theta ^-$. 
We introduce the fermionic derivatives w.r.t.~$ \theta ^\pm$, $D_\pm$, and those w.r.t.~$ \hat\theta ^\pm$, $\hat 
D_\pm$. They are defined by,
\begin{eqnarray}
D_+{}^2=\hat D_+{}^2= -\frac{i}{2} \,\partial _\pp\,, \quad
D_-{}^2=\hat D_-{}^2= -\frac{i}{2} \,\partial _=,
\end{eqnarray} 
and all other anti-commutators vanish.

The action $ {\cal S}$ in $N=(2,2)$ superspace is given by,
\begin{eqnarray}
{\cal S}=\int\, d{}^2\sigma\,d{}^2 \theta \,d{}^2 \hat \theta \, {\cal V}. 
\end{eqnarray}
As the measure has dimension zero, the Lagrange density $ {\cal V}$ can only be some function of scalar $N=(2,2)$ 
superfields. It is clear that in order to generate dynamics, one will have to judiciously constrain the 
$N=(2,2)$ superfields.

Consider a set of bosonic, scalar $N=(2,2)$ superfields $X^a$, $a\in\{1,\cdots D\}$. Expanding them in powers of
$\hat \theta^+$ and $\hat \theta ^-$, one finds that each general $N=(2,2)$ superfield consists of 
4 $N=(1,1)$ superfields. As a warming up exercise, let us first study those constraints linear in the derivatives that
reduce the number of $N=(1,1)$ components in a general $N=(2,2)$ superfield to one. 
Fixing both $\hat D_+X^a$ and $\hat D_-X^a$ simultaneously does the job. The most general constraints consistent 
with dimensions and Lorentz covariance are then given by, 
\begin{eqnarray}
\hat D_\pm X^a=J_\pm^a{}_{b}(X) D_\pm X^b,\label{pmcon}
\end{eqnarray}
where $J_\pm(X)$ are at this point two arbitrary $(1,1)$ tensors. 
From this, one gets immediately,
\begin{eqnarray}
\hat D_+^2X^a&=& +\frac{i}{2}(J_+^2)^a{}_b \partial_\pp\,X^b+ \frac 1 2 {\cal N}[J_+,J_+]^a{}_{bc} D_+X^b\, D_+X^c, 
\nonumber\\
\hat D_-^2X^a&=& +\frac{i}{2}(J_+^2)^a{}_b \partial_=\,X^b+\frac 1 2 {\cal N}[J_-,J_-]^a{}_{bc} D_-X^b\, D_-X^c,
\end{eqnarray}
and\footnote{Out of two {\em commuting} $(1,1)$ tensors $R^a{}_b$ and $S^a{}_b$, one contructs a $(1,2)$ tensor,
${\cal M}[R,S]^a{}_{bc}=(R^a{}_dS^d{}_{b,c}-
S^a{}_dR^d{}_{c,b}+S^d{}_bR^a{}_{c,d}-R^d{}_cS^a{}_{b,d})/2$. One has that $ {\cal M}[R,S]^a{}_{bc}=- 
{\cal M}[S,R]^a{}_{cb}$ and ${\cal N}[R,S]^a{}_{bc}=
{\cal M}[R,S]^a{}_{bc}+{\cal M}[S,R]^a{}_{bc}$.} 
\begin{eqnarray}
\{\hat D_+,\hat D_-\}X^a=[J_+,J_-]^a{}_b D_- D_+X^b+2 {\cal M}[J_-,J_+]^a{}_{bc} D_+X^b D_-X^c.
\end{eqnarray}
Requiring that these constraints are consistent with $\hat D_+^2= -(i/2) \partial_\pp\,$,
$\hat D_-^2= -(i/2) \partial_=$ and $\{\hat D_+,\hat D_-\}=0$, we get the following conditions,
\begin{eqnarray}
&&J_\pm^2=-{\bf 1},\quad {[}J_+,J_-{]}=0\quad , \nonumber\\
&&{\cal N}{[}J_\pm,J_\pm{]}= {\cal M}{[}J_+,J_-{]}=0.\label{bothchir}
\end{eqnarray}
All conditions in eq.~(\ref{bothchir}) are necessary. Indeed, in the literature it is often erroneously stated that 
two complex structures are simultaneously integrable if they commute\footnote{While this statement is made in numerous 
papers, we just mention one paper for which one of the present authors is responsible \cite{Sevrin:1996jr}. 
In that paper a multiplicative factor of two is lacking in front of the last term in eq.~(A.6), which invalidates the 
proof of lemma 1.}. 

In other words, 
$J^2_\pm=-{\bf 1}$, $ {\cal N}[J_\pm,J_\pm]=0$ and $[J_+,J_-]=0$ would imply that $ {\cal N}[J_+,J_-]=0$ holds. 
However this is wrong! A very nice and explicit counter example was provided in \cite{Gauntlett:2003cy}.
Consider a six-dimensional target manifold with the following two complex structures,
\begin{eqnarray}
J_+= \left(
\begin{array}{cccccc}
0 & -1&0 &0 &0 & 0 \\
1  & 0&0 &0 &0 &0  \\
0   &0 & 0&1 &0 &0  \\
  0  &0 & -1& 0&0 &0  \\
-X^1	 &-X^1 &X^3 &-X^3 &0 &-1  \\
-X^1	  &X^1 &X^3 &X^3 &1 & 0 \\
\end{array}
\right),
\end{eqnarray}
and
\begin{eqnarray}
J_-= \left(
\begin{array}{cccccc}
0 & -1&0 &0 &0 & 0 \\
1  & 0&0 &0 &0 &0  \\
0   &0 & 0&1 &0 &0  \\
  0  &0 & -1& 0&0 &0  \\
-X^1	 &X^1 &X^3 &X^3 &0 &1  \\
-X^1	  &-X^1 &X^3 &-X^3 &-1 & 0 \\
\end{array}
\right).
\end{eqnarray}
One readily verifies that they are complex structures, 
{\em i.e.} $J_\pm^2=-{\bf 1}$ and $ {\cal N}{[}J_\pm,J_\pm{]}=0$. 
They commute, ${[}J_+,J_-{]}=0$, as well. 
However, when calculating $ {\cal M}{[}J_+,J_-{]}$, one finds that all components vanish 
except for,
\begin{eqnarray}
&&{\cal M}{[}J_+,J_-{]}^5{}_{11}={\cal M}{[}J_+,J_-{]}^5{}_{21}={\cal M}{[}J_+,J_-{]}^5{}_{22}
={\cal M}{[}J_+,J_-{]}^5{}_{43}=
{\cal M}{[}J_+,J_-{]}^6{}_{21}= \nonumber\\
&&\qquad{\cal M}{[}J_+,J_-{]}^6{}_{33}={\cal M}{[}J_+,J_-{]}^6{}_{43}={\cal M}{[}J_+,J_-{]}^6{}_{44}=
+1, \nonumber\\
&&{\cal M}{[}J_+,J_-{]}^5{}_{12}={\cal M}{[}J_+,J_-{]}^5{}_{33}={\cal M}{[}J_+,J_-{]}^5{}_{34}
={\cal M}{[}J_+,J_-{]}^5{}_{44}= 
{\cal M}{[}J_+,J_-{]}^6{}_{11}= \nonumber\\
&&\qquad{\cal M}{[}J_+,J_-{]}^6{}_{12}={\cal M}{[}J_+,J_-{]}^6{}_{22}={\cal M}{[}J_+,J_-{]}^6{}_{34}=-1.
\end{eqnarray}
Consequently, the mixed Nijenhuis tensor $ {\cal N}[J_+,J_-]={\cal M}[J_+,J_-]+ {\cal M}[J_-,J_+]$ 
does not vanish! In other words, all four conditions in eq.~(\ref{bothchir}) have to be imposed independently. 
The non-linear $ \sigma $-model constructed out of superfields constrained as in eq.~(\ref{pmcon}) so that 
eq.~(\ref{bothchir}) holds, will be such that eqs.~(\ref{hermit}) and (\ref{covconst}) are automatically 
satisfied\footnote{One can also look at things from the point of view of the $N=(1,1)$ non-linear 
$ \sigma $-model without referring to $N=(2,2)$ superspace. 
Using the covariantly constancy of the complex structures, one shows $2\, {\cal M}^a{}_{bc}{[}J_+,J_-{]}=
{[}J_+,J_-{]}^a{}_d \Gamma ^d_{(-)}{}_{bc}$. Put differently: the fact that two complex structures commute and that 
they are covariantly constant does imply that they are simultaneously integrable. However, we reiterate that in an 
$N=(2,2)$ superspace treatment one 
only deals with eq.~(\ref{bothchir}) which will imply -- at least for the case of 
commuting complex structures -- eqs. (\ref{hermit}) and (\ref{covconst}).}.

Let us now come back to eq.~(\ref{bothchir}) which does imply that both $J_+$ and $J_-$ are simultaneously integrable. 
This means that we can make a coordinate transformation such that both $J_+$ and $J_-$ are diagonal with eigenvalues 
$\pm i$. If the eigenvalue of $J_+$ and $J_-$ have the same (opposite) sign, we are dealing with a (twisted) 
chiral field. Indeed, a chiral field $Z$, and its hermitean conjugate $ \bar Z$, satisfy,
\begin{eqnarray}
\hat D_\pm Z=+i\, D_\pm Z,\quad \hat D_\pm \bar Z=-i\, D_\pm \bar Z,
\end{eqnarray}
while for a twisted chiral field $Y$ (and its hermitean conjugate $ \bar Y$) we get,
\begin{eqnarray}
\hat D_+ Y= +i\, D_+Y,\quad \hat D_-Y=-i\,D_-Y,\quad
\hat D_+ \bar Y= -i\, D_+ \bar Y,\quad \hat D_- \bar Y=+i\,D_- \bar Y.
\end{eqnarray}
The first explicit example of a non-linear $ \sigma $-model which requires both chiral and twisted chiral superfields
-- the $S^3\times S^1$ WZW model -- was given in \cite{Rocek:1991vk}.

We now generalize the constraints by allowing for additional auxiliary fields. We start from a set of $m$ general 
$N=(2,2)$ superfields which we combine in an $m\times 1$ matrix $\IX$. The most general right handed constraint linear 
in the derivatives which we can impose is,
\begin{eqnarray}
\hat D_+\IX=\IJ_+\, D_+\IX+\IK_+ \,\Psi_+,\label{gp1}
\end{eqnarray}
where $\Psi_+$ is a $m\times 1$ column matrix of fermionic $N=(2,2)$ superfields and
$\IJ_+$ and $\IK_+$ are {\em constant} $m\times m$ matrices\footnote{In order to keep the analysis feasible, we only 
consider constraints linear in the superfields. This is a reasonable simplification as non-linear constraints would 
significantly complicate -- if not make it impossible -- the quantization of the resulting non-linear $ \sigma 
$-model.}. 
Through a linear transformation we can bring $\IK_+$ in its Jordan normal form. Making an appropriate linear 
combination of the components of $\Psi_+$ allows one to reduce $\IK_+$ to the form of a projection operator,
\begin{eqnarray}
\IK_+=\left(\begin{array}{cc}
{\bf 0}&{\bf 0}\\
{\bf 0}&{\bf 1}_{k\times k}
\end{array}
\right)\,,
\end{eqnarray}
where ${\bf 1}_{k\times k}$ is the $k\times k$ ($k\leq m$) unit matrix. It is clear that $k$ determines the number of 
fermionic (auxiliary) superfields which will appear in eq.~(\ref{gp1}). We now split $\IX$ into a $k\times 1$ matrix 
$\tilde X$ and an $(m-k)\times 1$ matrix $\check X$ of superfields. Appropriately redefining the fermionic superfields 
in $\Psi_+$, we can rewrite eq.~(\ref{gp1}) without any loss of generality as,
\begin{eqnarray}
\hat D_+\check X&=&J_+\,D_+\check X+ K_+\,D_+\tilde X,
\nonumber\\
\hat D_+\tilde X&=&\tilde\psi_+,\label{gp2}
\end{eqnarray} 
where $J_+$ and $K_+$ resp.~are $(m-k)\times(m-k)$ and $(m-k)\times k$ constant matrices resp.~and $\tilde\psi_+$ is a 
$k\times 1$ column matrix of fermionic (auxiliary) superfields. Implementing $\hat D_+{}^2=-(i/2) \partial _\pp\,$ in 
eq.~(\ref{gp2}) yields,
\begin{eqnarray}
K_+=0,\qquad J_+{}^2=-{\bf 1}_{(m-k)\times (m-k)}.
\end{eqnarray}
This implies that $m-k$ should be even and we put $2n=m-k$ with $n\in\IN$.
At this point we can, through an appropriate linear transformation on $\check X$, diagonalize $J_+$ with eigenvalues 
$\pm i$. Summarizing, we get -- without any loss of generality -- the following right handed constraints,
\begin{eqnarray}
\hat D_+\check X&=&i\,\check\IP\,D_+\check X,
\nonumber\\
\hat D_+\tilde X&=&\tilde\psi_+,\label{gp3}
\end{eqnarray}
where $\check\IP$ is given by,
\begin{eqnarray}
\check \IP=\left(\begin{array}{cc}
{\bf 1}_{n\times n}&{\bf 0}\\
{\bf 0}&-{\bf 1}_{n\times n}
\end{array}
\right)\,.
\end{eqnarray}
At this point we can still make arbitrary linear transformations on $\tilde X$. 
On $\check X$, linear transformations which commute with $\check\IP$ are allowed as well.
This freedom will be used later on.

The most general left handed constraints are of a form similar to the one in eq.~(\ref{gp1}), 
\begin{eqnarray}
\hat D_-\IX=\IJ_-\, D_-\IX+\IK_- \,\Psi_-.\label{gm1}
\end{eqnarray}
This can be further simplified by looking at the non-linear $ \sigma $-model we ultimately want to describe. 
We start with a Lagrange density $ {\cal V}(\IX)$ which is some scalar function of the superfields $\IX$. 
Integrating over $\hat \theta {}^\pm$, we get schematically the following dependence 
on $\Psi_+$ and $\Psi_-$ of the action in $N=(1,1)$ superspace,
\begin{eqnarray}
{\cal S}=\int d^2 \sigma d^2 \theta d^2 \hat\theta\, {\cal V}(\IX)=\int d^2 \sigma d^2 \theta 
\left(A_1+\Psi_+^TA_2+A_3\Psi_-+\Psi_+^TA_4\Psi_-\right),
\end{eqnarray}
where $A_\alpha$, $\alpha\in\{1,\cdots,4\}$ are matrices which depend on the remainder of the $N=(1,1)$ superfields. 
It is clear from this expression that the components of $\Psi_\pm$ appear as auxiliary fields. In order to solve for 
them, we have to require that the number of components of $\Psi_+$ which appear in eq.~(\ref{gp1}) equals the number 
of components of $\Psi_-$ appearing in eq.~(\ref{gm1}). 

A further exploration of the structure of the $N=(1,1)$ action will show that the right hand side 
of $\hat D_-\tilde X$ cannot 
contain any components of $\Psi_-$. Indeed, assume that some components of $\Psi_-$ do appear at the right
hand side of one or 
more components of $\hat D_-\tilde X$. Call these components $\tilde X^r$, $r\in\{1,\cdots ,\,l\leq k\}$. Making an 
appropriate 
field redefinition on those components of $\Psi_-$, we would get the following schematical structure for the 
constraints,
\begin{eqnarray}
\hat D_+\tilde X^r=\tilde \psi_+^r\,,\qquad
\hat D_-\tilde X^r=\tilde \psi_-^r\,.
\end{eqnarray}
Because of $\{ \hat D_+, \hat D_-\}=0$, we get,
\begin{eqnarray}
\hat D_+ \tilde \psi_-^r=-\hat D_- \tilde \psi_+^r\equiv F^r_{+-},
\end{eqnarray}
where when reducing to $N=(1,1)$ superspace, $F_{+-}^r$ will survive as new $N=(1,1)$ superfields. They will appear in 
the action as,
\begin{eqnarray}
{\cal S}=\int d^2 \sigma d^2 \theta d^2 \hat \theta  \, {\cal V}(\IX)=\int d^2 \sigma d^2 \theta \left(
\sum_{r=1}^l \partial _r\,{\cal V}(\IX)F^r_{+-}+\cdots
\right).
\end{eqnarray} 
The equations of motion for $F^r_{+-}$ force the potential ${\cal V}$ to be independent of $X^r$. Consequently we can 
impose that $ \hat D _-\tilde X$ does not depend on any of the components of $ \Psi_-$.

These considerations lead us to the conclusion that $2n\geq k$. Decomposing $\check X$ as a $k\times 1$ column matrix 
$X$ and a $2n-k$ column matrix $\hat X$, we arrive -- still without any loss of generality (but using input from the 
goal, the $ \sigma $-model, we aim for) -- at the following form for the constraints,
\begin{eqnarray}
&&\hat D_+X= i\,\IP\, D_+X,\quad \hat D_- X=\psi_-,\nonumber\\
&&\hat D_+\tilde X =\tilde\psi_+,\quad\hat D_-\tilde X= \tilde J_-\,D_-\tilde X+\tilde K_- \,D_-\hat X
+\tilde L_-\,D_-X\nonumber\\
&&\hat D_+\hat X=i\,\hat\IP \,D_+\hat X,\quad \hat D_-\hat X=\hat J_-\,D_-\hat X+\hat K_-\,D_-\tilde X
+\hat L_-\, D_-X+\hat M_-\,\psi_- ,\label{lefcon}
\end{eqnarray}
where $\psi_-$ is a $k\times 1$ column matrix of fermionic superfields which are related to the non-vanishing 
components of $\Psi_-$ through an appropriate coordinate transformation. Furthermore, we used the notation,
\begin{eqnarray}
\IP=\left(\begin{array}{cc}
{\bf 1}_{k/2\times k/2}&{\bf 0}\\
{\bf 0}&-{\bf 1}_{k/2\times k/2}
\end{array}
\right)\,,\quad
\hat \IP=\left(\begin{array}{cc}
{\bf 1}_{(n-k/2)\times (n-k/2)}&{\bf 0}\\
{\bf 0}&-{\bf 1}_{(n-k/2)\times (n-k/2)}
\end{array}
\right)\,,\label{defip}
\end{eqnarray}
where the reality properties of the fields force us to take
$k\in 2\,\IN$.

Rests us to impose the integrability conditions. One verifies that $0=\{\hat D_+,\hat D_-\}\tilde X$, 
$0=\{\hat D_+,\hat D_-\} X$ and $ \hat D_-{}^2X=-(i/2) \partial _=X$ resp.~give us expressions for 
$\hat D_-\tilde \psi_+$, $\hat D_+\tilde \psi_-$ and $\hat D_-\psi_-$.

Implementing the integrability conditions which follow from $0=\{\hat D_+,\hat D_-\} \hat X$, 
$ \hat D_-{}^2\tilde X=-(i/2) \partial _=\tilde X$ and $ \hat D_-{}^2\hat X=-(i/2) \partial _=\hat X$
reduce eq.~(\ref{lefcon}) to,
\begin{eqnarray}
&&\hat D_+X= i\,\IP\, D_+X,\quad \hat D_- X=\psi_-,\nonumber\\
&&\hat D_+\tilde X =\tilde\psi_+,\quad\hat D_-\tilde X= \tilde J_-\,D_-\tilde X+\tilde K_- \,D_-\hat X
-\tilde K_-\hat J_-\hat L_-\,D_-X,\nonumber\\
&&\hat D_+\hat X=i\hat\IP \,D_+\hat X,\quad \hat D_-\hat X=\hat J_-\,D_-\hat X
+\hat L_-\, D_-X+\hat J_-\hat L_-\,\psi_- ,\label{lbl}
\end{eqnarray}
with,
\begin{eqnarray}
&&\hat J_-{}^2=-{\bf 1}_{(2n-k)\times (2n-k)},\quad \tilde J_-{}^2=-{\bf 1}_{k\times k}, \nonumber\\
&&{[}\hat J_-, \hat \IP{]}=0, \nonumber\\
&& \hat L_-\IP=\hat \IP\hat L_-, \nonumber\\
&&\tilde J_-\tilde K_-=-\tilde K_-\hat J_-.
\end{eqnarray}
Combining the second of these equations with the freedom to make an arbitrary linear transformation on $\tilde X$ 
(making simultaneously the same transformation on $\tilde \psi _+$) allows one to put,
\begin{eqnarray}
\tilde J_-=i\,\IP,
\end{eqnarray}
where $\IP$ was defined in eq.~(\ref{defip}). A final simplification is achieved by making the following field 
redefinitions,
\begin{eqnarray}
\hat X&\rightarrow&\hat X'=\hat X-\hat J_-\hat L_-\,X, \nonumber\\
\tilde X &\rightarrow&\tilde X'=\tilde X+ \frac 1 2\tilde K_-\hat J_-\,\hat X+\frac 1 2 \tilde K_-\hat L_-\,X, 
\nonumber\\
\tilde\psi_+&\rightarrow&\tilde\psi'_+=\tilde \psi_++\frac i 2 \tilde K_-\hat J_- \hat\IP\, D_+\hat X'.
\end{eqnarray}
This reduces eq.~(\ref{lbl}) to,
\begin{eqnarray}
&&\hat D_+X= i\,\IP\, D_+X,\quad \hat D_- X=\psi_-,\nonumber\\
&&\hat D_+\tilde X' =\tilde\psi'_+,\quad\hat D_-\tilde X'= i\,\IP\,D_-\tilde X',\nonumber\\
&&\hat D_+\hat X'=i\hat\IP \,D_+\hat X',\quad \hat D_-\hat X'=\hat J_-\,D_-\hat X' ,
\end{eqnarray}
where,
\begin{eqnarray}
\hat J_-{}^2=-{\bf 1},\qquad {[}\hat J_-,\hat\IP{]}=0.
\end{eqnarray}
We still have the freedom to make a linear transformation on $\hat X'$ provided it commutes with $\hat\IP$. We use 
this freedom to diagonalize $\hat J_-$ with eigenvalues $\pm i$. Pending upon the sign of the eigenvalues, $\hat X'$ 
constitutes of chiral and twisted chiral superfields \cite{Gates:1984nk}. The remaining superfields $(X,\tilde X')$
are recognized as semi-chiral superfields \cite{Buscher:1987uw}.

While the analysis of \cite{Bredthauer:2005zx} raised the hope that other auxiliary field structures beyond 
semi-chiral superfields might exist, we showed in the present note that this is not so. We made two important 
restrictions: we limited ourselves to constraints which were both linear in the derivatives as well as linear in the 
superfields (and it might very well be that using an appropriate coordinate transformation any constraint non-linear 
in the superfields can be brought to a constraint linear in the fields). If one wants to quantize the model, the 
latter restriction is essentially unavoidable. This note 
strengthens the quite unique role of semi-chiral superfields. When studying $d=2$, $N=(2,2)$ non-linear 
$\sigma $-models in the presence of NS-NS fluxes, 
semi-chiral superfields will be very generic. Indeed, consider e.g.~a 
very large but quite simple class of integrable 
$\sigma$-models, the WZW models. In \cite{Spindel:1988nh} it was shown that any even-dimensional WZW model allowed for 
an $N=(2,2)$ supersymmetry. However, as argued in \cite{Rocek:1991vk}, only the $SU(2)\times U(1)$ 
model can be described
solely in terms of chiral and twisted chiral superfields. All other WZW models on even dimensional group manifolds 
will require semi-chiral superfields as well (some explicit examples can be found in \cite{Ivanov:1994ec} and 
\cite{Sevrin:1996jr}). Up till now, the literature on semi-chiral superfields is rather limited. We hope that the 
present note will raise the interest in them. 

\vspace{5mm}

\noindent {\bf Acknowledgments}:
We thank Jim Gates, Arjan Keurentjes, Ulf Lindstr\"om, Dario Martelli, Martin Ro\c cek and Jan Troost for useful 
comments, discussions 
and suggestions. This work was supported in part by the Belgian Federal Science Policy Office
through the Interuniversity Attraction Pole P5/27, in part by the European
Commission FP6 RTN programme MRTN-CT-2004-005104 and in part by the
``FWO-Vlaanderen'' through project G.0428.06.

\end{document}